\documentclass[a4paper, 12pt]{article}
\usepackage{a4}
\usepackage{amsfonts}
\usepackage{amssymb}
\usepackage{cite}
\usepackage{epsfig}
\usepackage{psfig}

\author{}

\newcommand{\ch}{{\rm ch}}
\newcommand{\be}{\begin{equation}}
\newcommand{\ee}{\end{equation}}
\newcommand{\bea}{\begin{eqnarray}}
\newcommand{\eea}{\end{eqnarray}}

\newcommand{\1}{{\bf 1}}
\newcommand{\3}{{\bf 3}}
\newcommand{\2}{{\bf 2}}
\newcommand{\4}{{\bf 4}}

\newcommand{\Anti}{{\bf Anti}}
\newcommand{\Sym}{{\bf Sym}}
\newcommand{\Adj}{{\bf Adj}}
\newcommand{\n}{{\bf n}}
\newcommand{\N}{{\bf N}}

\newcommand{\ov}{\overline}
\def\IR{\relax{\rm I\kern-.18em R}}

\def\IP{\relax{\rm I\kern-.18em P}}
\def\inbar{\vrule height1.5ex width.4pt depth0pt}
\def\IC{\relax\,\hbox{$\inbar\kern-.3em{\rm C}$}}

\def\K3{{\bf K3}}

\def\ov{\overline}

\begin{document}

\title{
\begin{flushright} \vspace{-2cm}
{\small MPP-2005-114} \end{flushright}
\vspace{4.0cm}
Non-Abelian Brane Worlds: \\
The Heterotic String Story
}
\vspace{1.0cm}
\author{\small Ralph~Blumenhagen, Gabriele Honecker and  Timo Weigand}
\date{}

\maketitle

\maketitle

\begin{center}
\emph{Max-Planck-Institut f\"ur Physik, F\"ohringer Ring 6, \\
  80805 M\"unchen, Germany } \\
\vspace{0.2cm}
\tt{blumenha,gabriele,weigand@mppmu.mpg.de}
\vspace{1.0cm}
\end{center}
\vspace{1.0cm}

\begin{abstract}
\noindent  
We discuss chiral supersymmetric compactifications  of  the
$SO(32)$ heterotic string on Calabi-Yau manifolds equipped with direct sums
of stable bundles with structure group $U(n)$. 
In addition we  allow for  non-perturbative heterotic five-branes.
These models are S-dual to Type I compactifications with
D9- and D5-branes, which by themselves are  mirror symmetric to general 
intersecting D6-brane models. 
For the construction of concrete examples we consider
elliptically fibered Calabi-Yau manifolds with $SU(n)$ bundles given by the spectral cover construction. The $U(n)$ bundles are obtained
via twisting by line bundles. We present a four-generation Pati-Salam and a three-generation
Standard-like model.

\end{abstract}

\thispagestyle{empty}
\clearpage

\tableofcontents

\section{Introduction}

Despite continuous studies of four-dimensional string compactifications it is believed
that only a very tiny subset of the vast landscape of all string vacua
has by now been addressed. So far most model building attempts  have focused on either the
$E_8\times E_8$ heterotic string or, more recently, on Type I strings
and its generalizations to Type II orientifolds. 

The extensively studied heterotic backgrounds include 
toroidal orbifolds with Wilson lines (see e.g.~\cite{Ibanez:1987sn}), Calabi-Yau manifolds 
given by hypersurfaces on toric varieties (see e.g.~\cite{Kreuzer:2000xy})
 and elliptically fibered 
Calabi-Yau manifolds equipped with $SU(n)$ bundles (see e.g.~\cite{Curio:1998vu,Andreas:1998ei,Braun:2005ux,Braun:2005bw})  defined
via the spectral cover construction of Friedman, Morgan and Witten
\cite{Friedman:1997yq}.

On the open string side  a promising class is given by intersecting D6-brane
models in Type IIA, which have primarily been studied on toroidal orbifolds
(see \cite{AU03,EK03,DL04,Blumenhagen:2005mu} for reviews).  
There have only been a few attempts to extend these
constructions to more generic smooth Calabi-Yau manifolds \cite{Blumenhagen:2002wn,Blumenhagen:2002vp}. 
Progress has mainly
been hampered by the fact that not very much is known about special Lagrangian
three-cycles on concrete Calabi-Yau's. Remarkably, the powerful techniques of 
boundary conformal field theory as applied in the framework of Gepner models allow for an algebraic construction of
a plethora of such orientifolds with the K\"ahler moduli fixed
at string scale radii \cite{Brunner:2004zd,bw04,Dijkstra:2004cc}. 

Via mirror symmetry geometric intersecting D6-brane models are related
to Type IIB orientifolds with D9(D7) and D5(D3) branes. As B-type branes
they also carry some in general non-abelian gauge bundles. S-duality~\cite{Polchinski:1995df} maps
such Type I models to compactifications of the $SO(32)$ heterotic
string.
 
A detailed study of this type of string vacua is of immediate interest. One
obvious reason is, of course,  the long-standing problem of implementing
Standard-model like physics into fully-fletched string compactifications. 
In view of the recent discussion about the landscape \cite{MD03},
one can formulate this question as: Does the MSSM or any generalisation
to be found at the LHC couple consistently to gravity such
that it can be realized as a compactification of string or
M-theory? In short:  Does it belong to the landscape or the swampland of 
low energy actions \cite{Vafa:2005ui}? 

As we will demonstrate, the $SO(32)$ heterotic string does provide us with a huge new class of quite promising vacua, which by S-duality significantly extends the set of open string models studied so far. 
Besides these more phenomenological questions, following the general spirit of the string landscape idea, it is desirable to compare and classify vacua in as different regions of the string moduli space as possible.    

In \cite{Blumenhagen:2005ga,Blumenhagen:2005pm} we have given a general description of $E_8\times
E_8$ and $SO(32)$ heterotic string 
compactifications on Calabi-Yau manifolds with general stable $U(n)$ vector
bundles. 
In particular, we have worked out the generalized Green-Schwarz
mechanism involving not only the universal axion but also the axio-K\"ahler fields. 
In addition we have computed the perturbatively exact one-loop correction to the
Donaldson-Uhlenbeck-Yau (DUY) equation, which turned out to be S-dual to
the perturbative part of the $\Pi$-stability condition of B-type  branes in Type IIB
string theory. 

Let us emphasise that, on the one hand, for the $E_8\times E_8$ string $SU(n)$ bundles\footnote{For some  
earlier works of compactifications with $U(n)$ bundles to four and six dimensions see 
also~\cite{Witten:1984dg,Distler:1987ee,Sharpe:1996xn,Lukas:1999nh,Andreas:2004ja}
and~\cite{Green:1984bx,Strominger:1986uh,Berglund:1998va,Aldazabal:1996du}, respectively.} are
the most natural objects to work with; the point is that they give nice breaking patterns of
$E_8$ down to some GUT gauge group and that for vanishing first Chern class
of the  bundle the one-loop corrected DUY equation is trivially satisfied. 
On the other hand for the $SO(32)$ heterotic strings $U(n)$ bundles
play an analogous role for the same two reasons. First there are natural embeddings
of $U(n)$ into $SO(32)$ and second for $U(n)$ there
is a  chance to compensate the generically non-vanishing one-loop correction
to the supersymmetry condition  by the tree-level part.
In \cite{Blumenhagen:2005pm} we mainly concentrated on the general framework of such 
$SO(32)$ compactifications
but only provided a small number of simple non-realistic examples. 
It is the aim of this paper to explore much more extensively the model building possibilities
and to give a very concrete set of models. 

To begin with we generalise 
the framework of \cite{Blumenhagen:2005pm} by including more general
$SO(32)$ breaking patterns which give rise to, for instance, chiral fields
transforming in the symmetric representation of the unitary gauge groups. 
In addition, we also consider contributions from non-perturbative heterotic
five-branes; in contrast to $E_8\times E_8$ five-branes \cite{Lukas:1998hk}, they  give
rise to chiral matter and anomaly cancelling Green-Schwarz couplings. 

Since for stability reasons we require 
supersymmetry at the string scale, a very explicit  description of
stable $U(n)$ bundles over some Calabi-Yau manifolds is needed. 
We point out that by a twisting procedure one can define stable $U(n)$ bundles
on elliptically fibered Calabi-Yau's starting with the stable $SU(n)$ bundles
of Friedman, Morgan and Witten \cite{Friedman:1997yq}.
Even though  in the course of this paper we work in the heterotic framework, 
the models we study can easily be S-dualized to compactifications of the Type I string
on smooth elliptically fibered Calabi-Yau's with stacks of D9-branes carrying non-trivial
$U(n)$ bundles as well as D5-branes\cite{bhwnew}. 

Using this large class of bundles we search for Standard-like
and GUT-like models, where the pattern for the chiral massless sectors is
similar to the intersecting D-brane models studied earlier. 

The remainder of this paper is organized as follows: Section~\ref{SecModel} gives the model building 
rules for $SO(32)$ heterotic string  compactifications on Calabi-Yau 
mani\-folds with products of $U(n)$ bundles and additional   
five-branes. In section~\ref{secUnbundles},
we review the spectral cover construction for $SU(n)$ bundles on 
elliptically fibered 
Calabi-Yau manifolds and introduce $U(n)$ bundles by twisting with line 
bundles.
For the base of the elliptic fibration chosen to be some del Pezzo
surface, in section~\ref{secdelPezzo} we present a four-generation Pati-Salam type
model and a three-generation MSSM like model. These are only two examples among a much larger class 
of models which our very restricted and preliminary search has produced. 
Section~\ref{secConclusions}
contains our conclusions. Finally, some technical details on the K\"ahler cone and 
anomaly cancellation are collected in appendices~\ref{Kcone} and~\ref{AppTrace}.

\section{The $SO(32)$ heterotic string with  unitary \mbox{bundles} and five-branes}
\label{SecModel}

Following our recent work \cite{Blumenhagen:2005pm}, in this section we review and extend
a general class of compactifications
of the $SO(32)$ heterotic string  involving 
direct sums of bundles with unitary structure groups.
This also includes direct sums of abelian bundles.
 
\subsection{Definition of a class of $SO(32)$ heterotic string vacua}

Concretely, we compactify the $SO(32)$ heterotic string on a Calabi-Yau manifold
$X$ and consider decompositions of the gauge group $SO(32)$ into 
\bea
SO(2M) \times \prod_{i=1}^{K} U(M_i)
\eea
with \mbox{$M+\sum_{i=1}^{K} M_i=16$}.
To achieve such a breaking of the original $SO(32)$ we give a background
values to gauge field strength of vector bundles  on the internal 
manifold with structure group   
\mbox{$G=\prod_{i=1}^{K} U(n_i)$}, 
where each $U(n_i)$ is diagonally embedded into a $U(M_i)$ gauge factor
with $M_i=n_i\, N_i$. 
Thus we  consider bundles of the form
\bea
\label{decomp1}
W = \bigoplus_{i=1}^{K} V_{i}
\eea
with $V_i$ denoting a  rank $n_i$ unitary bundle. 
The resulting observable non-abelian gauge group is
\bea 
H=SO(2M) \times \prod_{i=1}^K U(N_i),
\eea
where as usual maximally only
the anomaly-free part of the $U(1)^{K}$ gauge factors 
remains in the low energy gauge group.

In addition to this perturbative sector we  take into account the possible
contribution from heterotic 5-branes (H5-branes), which can be understood as
instantons of zero size \cite{Strominger:1990et,Duff:1990ya, Duff:1990wu,Duff:1991sz,Witten:1995gx}.
For supersymmetry each  H5-brane has to wrap an (in general reducible) holomorphic curve $\Gamma$ on $X$. This means that the associated class $[\Gamma]$ is effective,  i.e. lies inside the Mori cone of $X$. If $\Gamma$ is irreducible, this really corresponds to a single H5-brane and gives rise to an additional $Sp$(2) gauge group in the effective action. Otherwise one decomposes $\Gamma$ into the irreducible generators of the Mori cone $\Gamma_a$, $\Gamma  = \sum_a N_a \,\Gamma_a , \, N_a \in {\mathbb Z}_0^{+}$. Due to the multiple wrapping around each irreducible curve $\Gamma_a$, the additional gauge group in the effective action gets enhanced to $\prod_a Sp(2 N_a)$.
The decomposition into generators 
may not be unique and the gauge group may therefore vary in the different regions of the associated moduli space. However, its total rank and the total number of chiral degrees of freedom charged under the symplectic groups are only dependent on $\Gamma$, of course.

On the worldvolume of a  $SO(32)$ H5-brane lives a massless 
gauge field, which compared to the $E_8\times E_8$ H5-brane leads to  
different low energy physics.  Recall that in the latter case, the H5-brane supports
an antisymmetric self-dual 2-form field and the rank of the gauge factor depends on the genus of the wrapped two-cycle \cite{Lukas:1998hk}. 

In order to cancel the gravitational anomalies on the $SO(32)$ H5-brane 
world-volume and by heterotic-Type I duality, one can infer that  the effective low energy action on the H5-branes
has to have the usual Chern-Simons form
\bea
\label{CS}
S_{H5_a}= - \mu_5 \int_{\mathbb R_{1,3}\times \Gamma_a} \sum_{n=0}^1
B^{4n+2}\wedge \left( N_a+ {\ell_s^4\over 4(2\pi)^2} {\rm tr}_{Sp(2N_a)} F_a^2
\right) \wedge 
        { \sqrt{\hat{\cal A}({\rm T}\Gamma_a)}\over \sqrt{\hat{\cal A}({\rm N}\Gamma_a)} },
\eea
with the H5-brane tension  $\mu_5= \frac{1}{(2 \pi)^5 \,(\alpha')^3}$.
T$\Gamma_a$ and  N$\Gamma_a$ denote the tangent bundle and  the normal
bundle,  respectively, of the 2-cycle $\Gamma_a$, which  for concreteness we take to be irreducible from now on and wrapped by a stack of $N_a$ H5-branes. 
The curvature occurring in the definition of the $\hat{\cal A}$ genus is
defined as ${\cal R} = -i\ell_s^2 R$ ($\ell_s \equiv 2 \pi \sqrt{\alpha'}$). The sign of the Chern-Simons action is dictated by supersymmetry. With the choice in (\ref{CS}) the real part of the gauge kinetic function for the $Sp(2N_a)$-group is indeed positive, as we demonstrate in Appendix B.
Note that~(\ref{CS}) implies both the usual magnetic coupling to $B^{(6)}$ defined by $\ast_{10} dB^{(2)} = e^{2 \phi_{10}} d B^{(6)}$ 
and a coupling to $B^{(2)}$, which is essential to cancel the  mixed abelian-gravitational and abelian-symplectic
anomalies by a generalized Green-Schwarz mechanism 
(see Appendix B).

For our purpose it is useful to describe the H5-brane wrapping $\Gamma_a$ as 
 the skyscraper sheaf ${\cal O}\vert_{\Gamma_a}$ with the locally free
resolution given by   the Koszul sequence
\bea
 0\rightarrow \wedge^2 E_a^* \rightarrow E_a^* \rightarrow {\cal O}_X \rightarrow {\cal O}\vert_{\Gamma_a}\rightarrow 0.
\eea
Here $E_a$ is a vector bundle of rank two and  $\Gamma_a$ is defined as the zero locus
of a section of $E_a$.
The Chern characters of this sheaf can be computed to be $\ch({\cal
  O}\vert_{\Gamma_a})=(0,0,\gamma_a,0)$,
where $\gamma_a$ denotes the Poincar\'e dual four-form of the 2-cycle $\Gamma_a$.
However, due to the overall minus sign in (\ref{CS}) relative to the gauge bundles
on the Calabi-Yau, the five-brane Chern characters get an additional minus sign so that
in the following $\ch({\cal O}\vert_{\Gamma_a})=(0,0,-\gamma_a,0)$.

\subsection{The massless spectrum}

For the low energy effective theory it is important to know the massless
four-dimensional spectrum. 
The perturbative spectrum can be determined from the
 decomposition of the adjoint representation of $SO(32)$  
into representations of $\prod_i U(N_i)\times U(n_i)$
\bea
{\bf 496} \to
\left(\begin{array}{c} 
(\Anti_{SO(2M)})_0\\
\sum_{j=1}^{K} (\Adj_{U(N_j)};\Adj_{U(n_j)})\\
\sum_{j=1}^{K} (\Anti_{U(N_j)};\Sym_{U(n_j)}) + 
               (\Sym_{U(N_j)};\Anti_{U(n_j)}) + h.c.\\
\sum_{i < j} (\N_i,\N_j;\n_i,\n_j) + (\N_i,\ov{\N}_j,\n_i,\ov{\n}_j) + h.c. \\
\sum_{j=1}^{K} (2M, \N_j;\n_j) + h.c.\\
\end{array}\right).   
\eea

\begin{table}[htb]
\renewcommand{\arraystretch}{1.5}
\begin{center}
\begin{tabular}{|c||c|}
\hline
\hline
reps. & $H=\prod_{i=1}^K SU(N_i)\times U(1)_i \times SO(2M)\times \prod_{a=1}^L Sp(2N_a)$   \\
\hline \hline
$(\Adj_{U(N_i)})_{0(i)}$ & $H^*(X,V_i \otimes V_i^{\ast})$  \\
\hline
$(\Sym_{U(N_i)})_{2(i)}$ & $H^*(X,\bigwedge^2 V_i)$  \\
$(\Anti_{U(N_i)})_{2(i)}$ & $H^*(X, \bigotimes^2_s  V_i)$  \\
\hline
$(\N_i,\N_j)_{1(i),1(j)}$ & $H^*(X, V_i \otimes V_j)$ \\
$(\N_i,\ov \N_j)_{1(i),-1(j)} $ &  $H^*(X, V_i \otimes V_j^{\ast})$ \\
\hline
$(\Adj_{SO(2M)})$ & $H^*(X,{\cal O})$ \\
$(2{\bf M}, \N_i)_{1(i)}$ & $H^*(X, V_i)$\\
\hline
$(\Anti_{Sp(2N_a)})$ &  Ext$_X^*({\cal O}\vert_{\Gamma_a},{\cal O}\vert_{\Gamma_a})$ \\ 
$(\N_i,2\N_a)_{1(i)}$ & Ext$_X^*(V_i,{\cal O}\vert_{\Gamma_a})$  \\
$(2\N_a,2\N_b)$ & Ext$_X^*({\cal O}\vert_{\Gamma_a},{\cal O}\vert_{\Gamma_b})$  \\
\hline
\end{tabular}
\caption{\small Massless spectrum with 
the structure group  taken to be 
$G=\prod_{i=1}^{K} U(n_i)$.  }
\label{Tchiral1}
\end{center}
\end{table}
Due to the more general decomposition
of $SO(32)$, in contrast to \cite{Blumenhagen:2005pm}, we now also get chiral matter
in the symmetric representation of $U(N_i)$. In addition new chiral matter appears
from the non-perturbative H5-branes, which is absent for the H5-branes in $E_8\times E_8$ heterotic
string compactifications \cite{Lukas:1998hk}. In the latter case this is in accord
with the possibility of moving the 5-branes into the eleven-dimensional bulk
in the Horava-Witten theory. 
The resulting massless spectrum of the $SO(32)$ string arising both in the  perturbative and the
non-perturbative sector  is listed in Table~\ref{Tchiral1}.  

The matter arising in the H5-brane sector is described by appropriate extension groups. Following for instance \cite{Katz:2002jh}, 
the global extension groups Ext$^*_X ({\cal E},{\cal F})$
of two coherent sheaves on $X$ give the sheaf theoretic generalisation of the
cohomology groups $H^*(X,{\cal E}\otimes {\cal F}^*)$ for vector bundles on smooth manifolds. 
In particular, one can show that \cite{Katz:2002jh} 
\bea
{\rm Ext}_X^1 ({\cal O}\vert_{\Gamma_a},{\cal O}\vert_{\Gamma_a})=H^1(\Gamma_a,{\cal O})+H^0(\Gamma_a,{\rm N}\Gamma_a),
\eea
where the first term contains the possible Wilson line moduli  on the H5-brane
and the second term the geometric deformations of the two-cycles $\Gamma_a\subset X$. 
All these chiral supermultiplets transform in the antisymmetric representation of the symplectic 
gauge factor.

The perturbative chiral spectrum can be determined from the Euler characteristics of the various 
bundles $U$ occurring in the decomposition of $SO(32)$,
\bea 
\label{RRH}
         \chi(X,U)=\sum_{n=0}^3  (-1)^n \, {\rm dim}\, H^n(X,U)
         =\int_{X}\left[ {\rm ch}_3(U)+
           {1\over 12}\, c_2(T)\, c_1(U) \right].
\eea
For more general coherent sheaves the righthand side in (\ref{RRH}) measures
the alternating sum of the dimensions of the global extensions. 
Note that in the non-perturbative sector, the H5-branes give rise to chiral
matter in the bifundamental $(\N_i, 2\N_a)_{1(i)}$, which is counted by the index
\bea
\label{ind}
\chi(X, V_i{\otimes \cal O}\vert^*_{\Gamma_a} )  = -\int_X c_1(V_i) \wedge \gamma_a.
\eea
Again $\gamma_a$ denotes the Poincar\'e dual 4-form corresponding to the 2-cycle $\Gamma_a$.
Consistent with the absence of chiral matter for  symplectic gauge groups only, 
for two H5-branes wrapping 2-cycles $\Gamma_a$ and $\Gamma_b$ one gets
$\chi(X, {\cal O}\vert_{\Gamma_a}  \otimes {\cal O}\vert^*_{\Gamma_b})=0$.

\subsection{Model building constraints}

Let us summarize   the  
major model building constraints arising
from supersymmetry and tadpole cancellation \cite{Blumenhagen:2005pm}
including also H5-branes.

\begin{itemize}

\item
The vector bundles have to be holomorphic and 
the non-holomorphic Hermitian  Yang-Mills equation is modified at one-loop. 
The perturbatively exact integrability condition reads
\bea
\label{DUYloop}
{1 \over 2}    \int_X J\wedge J \wedge c_1(V)  - 
    g_s^2\, \ell_s^4 \, \int_X 
\left( {\rm ch}_3 (V)+\frac{1}{24}\, c_1(V)\,  c_2(T)\right)=0,
\eea
where $g_s=e^{\phi_{10}}$ and $\ell_s=2\pi\sqrt{\alpha'}$. Of course this equation has to hold inside the K\"ahler cone.
Also the local supersymmetry variation of the gauge fields
change accordingly and a unique supersymmetric solution exists if the
bundle is $\pi_h$-stable (which is the S-dual of the perturbative part
of  Type IIB $\Pi$-stability \cite{Douglas:2000ah}). In fact, one can show that any $\mu$-stable bundle 
is also
$\pi_h$-stable \cite{Enger:2003ue}. Strictly speaking, this is true in the
    perturbative regime, i.e. for sufficiently large  $r_i \gg 1$ and
small $g_s \ll 1$.
This is the regime  in which we work anyway 
to suppress non-perturbative corrections.

An additional constraint arises from the requirement 
that the one-loop corrected gauge couplings are real
\bea
\label{Gauged}
    {n\over 3!}\,
 \int_X J \wedge J \wedge J -
g_s^2\, \ell_s^4\,  \int_X J \wedge 
\left( {\rm ch}_2 (V)+\frac{n}{24}\, c_2(T)\right) >0
\eea
in the perturbative regime, i.e. for $g_s \ll 1$ and $r_i \gg 1$. 
Note that there will be additional  non-perturbative
stringy and $\alpha'$ corrections
to (\ref{DUYloop}) and (\ref{Gauged}).

\item
The Bianchi identity for the three-form field strength $H_3$ is given by
\bea
dH_3={\ell_s^2}\left( {1\over4\,  (2\pi)^2}\left[ 
{\rm tr} R^2 -{\rm tr} F^2 \right]+ \sum_a N_a \,\gamma_a   \right),
\eea
where the traces are taken in the fundamental representation of $SO(1,9)$ and $SO(32)$, respectively.
We have included the  magnetic coupling of the H5-branes.
For decompositions of the type~(\ref{decomp1}), 
the resulting constraint from integrating the perturbative and H5-brane contributions over an 
internal 4-cycle reads
\bea \label{TP}
\sum_{i=1}^{K} N_i \; \ch_2(V_i)  - \sum_{a=1}^L N_a \gamma_a= -c_2(T)  
\eea
in cohomology.
\item
The absence of a global Witten anomaly on the H5-branes requires that the number of chiral fermions in the fundamental of the 
$Sp(2 N_a)$ groups be even \cite{Witten:1982fp}.
In order to satisfy this constraint not only in the vacuum, but also in every topological sector of the theory, we insist 
that this holds even for every probe brane \cite{Uranga:2000xp}.
Inspection of (\ref{ind}) translates this into the constraint
\bea
\label{K}
c_1(W) =\sum_i N_i\, c_1(V_i) \in H^2(X, 2{\mathbb Z}).
\eea
Note that  this condition was originally derived
from the absence of anomalies in  the two-dimensional non-linear
sigma model~\cite{Witten:1985mj,Freed:1986zx}. 
In Type I string theory this is nothing else than the torsion K-theory
constraint for the non-BPS D7-brane \cite{Uranga:2000xp}.  
\end{itemize}

\section{Construction of U(n) bundles}
\label{secUnbundles}

We consider elliptically fibered Calabi-Yau manifolds with a section
and define  $SU(n)$ vector bundles \`{a} la Friedman, Morgan and Witten 
(FMW) via the  spectral
cover construction \cite{Friedman:1997yq}. This construction has originally been designed
for vector bundles with $SU(n)$ structure group and has been generalized
to $U(n)$ bundles in \cite{Andreas:2004ja}.
Here we take a different approach and simply define 
$U(n)$ bundles by twisting $SU(n)$ bundles with line bundles. 
But first let us briefly review the  spectral cover construction.

\subsection{Elliptically fibered  Calabi-Yau manifolds}

An elliptically fibered Calabi-Yau threefold $X$ consists of a 
complex two-dimensional base $B$ and a projection
\bea
       \pi:X\to B
\eea
with the property that the fiber over each point is an elliptic curve $E_b$.
One  requires that these bundles admit a section
\bea
    \sigma:B\to X
\eea
which identifies the base as a submanifold of the space $X$.
One assumes that the elliptic fibers are described by a 
Weierstrass equation,
\bea
       zy^2=4x^3-g_2 xz^2-g_3 z^3
\eea
where $g_2$ and $g_3$ are sections of the line bundles ${\cal L}^4$
and ${\cal L}^6$ respectively. For $X$ to be a Calabi-Yau manifold
the canonical line bundle ${\cal K}_X$ has to be trivial, which
in this case tells us that ${\cal L}={\cal K}_B^{-1}$, i.e.
\bea
     c_1({\cal L})=c_1({\cal K}_B^{-1})=c_1(B) .
\eea
If the base is smooth and preserves only ${\cal N}=1$ supersymmetry
in four dimensions, it is restricted to a del Pezzo surface, a Hirzebruch
surface, an Enriques surface or a blow up of a Hirzebruch surface.
In \cite{Friedman:1997yq} it has been shown that the second Chern class of the tangent
bundle can be expressed as
\bea
    c_2(TX)=12\sigma c_1(B) + c_2(B) +11c_1(B)^2,
\eea
where $\sigma$ denotes the two-form of the section, i.e. the Poincar\'{e} dual of
the four cycle $B\subset X$. It has also been shown that
$\sigma^2=-\sigma c_1(B)$.

\subsection{The spectral cover construction}

A set  of vector bundles with structure group $SU(n)$ on $X$ can be described
by the so-called spectral cover construction. 
The idea is to use a simple description of $SU(n)$ bundles over
the elliptic fibers and then globally glue them together to define
bundles over $X$. One of the many nice properties of such a construction is
that eventually the Chern classes of such bundles can be computed 
entirely in terms of data defined on
the base $B$. 

The starting point is that on the elliptic fibers semi-stable $SU(n)$ vector bundles
split into direct sums of line bundles of degree zero. 
That means that each line bundle can be written as ${\cal O}(Q_i)\otimes {\cal O}(p)^{-1}$,
where $Q_i$ is a unique point on the elliptic fiber. 
Therefore, for an $SU(n)$ bundle one obtains an $n$-tuple of such points fibered over
the base of the elliptic fibration. This is the so-called spectral cover $C$, which
is a ramified  $n$-fold cover of $B$ with $\pi_C:C\to B$. 
In order to specify the cohomology class $[C]$ of $C$ in $H^2(X,\mathbb Z)$, one has the freedom to twist
by an additional line bundle $\eta$ whose restriction
to the elliptic fiber is of degree zero.
For $SU(n)$ bundles this restriction has to be
the trivial line bundle on the fiber, which can be achieved by choosing
for $\eta$  the pull-back of a line bundle  on the base. Therefore, 
the spectral cover lies in the cohomology class
\bea
      [C]=n\sigma + \eta
\eea
with $\eta$ being an effective class in $B$.

In order to specify a bundle $V$ on $X$, besides the spectral cover $C$ in addition 
one has to choose a line bundle ${\cal N}$ on $C$ such that $V\vert_B=\pi{_C*}{\cal N}$,
where the push-forward bundle $\pi{_C*}{\cal N}$ is a bundle on $B$ whose
fiber at a point $b$ is the direct sum of $n$ lines ${\cal N}\vert_{Q_i}$. 

The entire vector bundle $V$ on $X$ is described by using the Poincar\'{e} bundle ${\cal P}$,
which is a line bundle on the fiber product $Y=X\times_B C$ with
\bea
   c_1({\cal P})=\Delta-\sigma_1-\sigma_2-c_1(B).
\eea
Here $\Delta$ denotes the diagonal divisor in $Y$ and
$\sigma_1$ and $\sigma_2$ the sections $\sigma_1:B\to X\subset Y$ and
$\sigma_2:B\to C\subset Y$. 
The vector bundle $V$ is then defined as
\bea
     V=\pi_{1*}(\pi_2^*\, {\cal N}\otimes {\cal P}),
\eea
where $\pi_1$ and $\pi_2$ denote the two projections of the fiber product $X\times_B C$ onto
the two factors $X$ and $C$.  

For $SU(n)$ bundles the condition $c_1(V)=0$ allows
one to fix $c_1({\cal N})$ in terms of $\sigma,\eta, c_1(B)$ and a rational number
$\lambda$
\bea
    c_1(V)=\pi_{C*}\left(  c_1({\cal N}) +{1\over 2} c_1(C) - {1\over 2} \pi^*_C\,  c_1(B)-\gamma\right)
\eea
with $\gamma=\lambda(n\sigma -\pi^*_C\eta +n\pi^*_C \, c_1(B))$.
The other two Chern classes  $c_2(V)$ and $c_3(V)$ 
can be determined by applying the Grothendieck-Riemann-Roch (GRR) theorem 
\bea
\label{GRR}
       \pi_{1*}\biggl( {\rm ch}(\pi_2^*\, {\cal N}\otimes {\cal P}) \, 
         {\rm Td}(X\times_B C)\biggr)=         {\rm ch(V)}\, {\rm Td}(X),
\eea
where the push-forward $\pi_{1*}$ of a  form is 
defined as integration over the fiber.
For completeness let us list the resulting Chern classes:
\begin{itemize}
\item{ The first Chern class vanishes
\bea
   c_1(V)=0.
\eea
}
\item{
The second Chern class can be expanded in a piece along the base and a piece
along the fiber
\bea
     c_2(V)=\eta\, \sigma + \omega
\eea
with $\omega$ defined as 
\bea
   \omega=-{1\over 24}\, c_1(B)^2\, (n^3-n) +{1\over 2} \left( \lambda^2 -{1\over 4}\right) n\eta\,
   (\eta-nc_1(B)).
\eea
}
\item{
The third Chern class after integration over the fiber has the following simple
form
\bea  
   c_3(V)=2\,\lambda \eta(\eta-nc_1(B))
\eea
and leads to the Euler characteristic
\bea
  \chi(V)=\lambda \eta(\eta-nc_1(B)).
\eea
}
\end{itemize} 
Note that all classes can be expressed essentially by classes 
living on the base manifold
$B$. 
These formulas are supplemented by additional  relations for the defining quantities
guaranteeing  that $c_1({\cal N})$ is an integer class:
\bea
  && n\ {\rm odd}: \phantom{i}\lambda\in \mathbb Z+{1\over 2}  \\
  && n\ {\rm even}: \lambda\in \mathbb Z, \phantom{aaai} \quad \eta+c_1(B)=0\ {\rm mod}\ 2 \\
  && \phantom{in\ {\rm even}:}  \lambda\in \mathbb Z+{1\over 2}, \quad c_1(B)=0\ {\rm mod}\ 2. 
\eea

The bundles of this construction so far are $\mu$-semi-stable. For them 
to be truly stable the spectral cover needs to be irreducible \cite{Friedman:1997ih}. 
It has been shown in \cite{Donagi:2004ia} 
that this boils down to two conditions
on the curve $\eta$:
\begin{itemize}
\item{ The linear system $|\eta|$ has to be base point free}
\item{ The curve $\eta-n\, c_1(B)$ has to be effective.}
\end{itemize}
If this is the case then the $SU(n)$ bundles are $\mu$-stable.\footnote{In
  fact, the proof of stability assumes that the K\"ahler parameter of the
  fiber lies in a certain range near the boundary of the 
  K\"ahler cone, i.e $J=\epsilon \sigma +J_B$ with
sufficiently small $\epsilon$ \cite{Friedman:1997ih}. Since the value of $\epsilon$ is not
known, in all models involving the spectral cover constructions it is
  therefore a subtle issue if the region of stability overlaps with the
  perturbative regime, which is needed to have control over non-perturbative
  effects. We thank B. Andreas for drawing our attention to this issue.}

\subsection{Twisting by a line bundle}

One way to define stable $U(n)$ bundles is via twisting with 
a line bundle on $X$. 
As before we start with a stable $SU(n)$ bundle as it
arises  from the Fourier Mukai transform
\bea
     V=\pi_{1*}(\pi_2^*\, {\cal N}\otimes {\cal P})
\eea
of  any line bundle ${\cal N}$ over the spectral cover $C$.
In addition  we take an arbitrary line bundle ${\cal Q}$ on $X$ 
with
\bea
     c_1({\cal Q})=q\, \sigma + c_1(\zeta)
\eea
and $c_1(\zeta)\in H^2(B,\mathbb Z)$. 
Then we can define the twisted bundle
\bea
      V_{{\cal Q}}=V\otimes {\cal Q},
\eea
which has non-vanishing first Chern class  unless ${\cal Q}$ is trivial.
A bundle $V$ is $\mu$-stable if
and only if $V\otimes {\cal Q}$ is stable for every line bundle ${\cal Q}$.\cite{Mumford,Takemoto}
Therefore, all these twisted $U(n)$ bundles are $\mu$-stable
if the $SU(n)$ bundle is.

Using ${\rm ch}(V\otimes {\cal Q})={\rm ch}(V)\,  {\rm ch}({\cal Q})$
it is an easy exercise to  compute the resulting Chern classes: 

\begin{itemize}
\item{ The first Chern class is given by
\bea
   c_1(V)=n\, q\,\sigma+{n}\, c_1(\zeta).
\eea
}
\item{
The second Chern character can be expanded in a piece along the base and a piece
along the fiber
\bea
     {\rm ch}_2(V)=\left[ -\eta+{n\, q\over 2} \biggl(2\,c_1(\zeta)-q\,
       c_1(B)\biggr)\right] \sigma + a_F
\eea
with 
\bea
     a_F={n\over 2} c_1(\zeta)^2 -\omega
\eea
and $\omega$ defined as usual
\bea
   \omega=-{1\over 24}\, c_1(B)^2\, (n^3-n) +{1\over 2} \left( \lambda^2 -{1\over 4}\right) n\eta\,
   (\eta-nc_1(B)).
\eea
}
\item{
The third Chern character after integration over the fiber has the following form
\bea  
   {\rm ch}_3(V)&=&\lambda \eta(\eta-nc_1(B)) -  \eta\, c_1(\zeta) +
    q\left({n\over 2}\, c_1(\zeta)^2 -\omega\right) + \nonumber\\
   && q\, c_1(B)\left(
    \eta-{n\, q\over 2}\, c_1(\zeta)  +{n\, q^2\over 6}\, c_1(B) \right),
\eea
which leads to the Euler characteristic
\bea
  \chi(V)&=&\lambda \eta(\eta-nc_1(B)) - (\eta-n c_1(B))\,
  c_1(\zeta)+  q\left({n\over 2}\, c_1(\zeta)^2 -\omega\right) + \\
   &&q\, c_1(B)\left(
    \eta-{n\, q\over 2}\, c_1(\zeta)  +{n\, q^2\over 6}\, c_1(B) \right) +
    {n\, q\over 12}\left( c_2(B) - c_1(B)^2\right)  .\nonumber
\eea
}
\end{itemize}
Such twisted unitary bundles will be the class of bundles we are 
going to use for our search of Standard-like and GUT models.
The above construction also includes $U(1)$ bundles, if $V$ is simply the 
trivial line bundle. In the formulas for the Chern classes this boils
down to setting $n=1$ and $\eta=0$.

In order to compute the Euler characteristics of products of bundles $V_i\otimes V_j$
appearing in Table (\ref{Tchiral1}) one utilizes the formula
\bea
   \chi(V_i\otimes V_j)=n_i\, \chi(V_j) + n_j\, \chi(V_i) + c_1(V_i)\, {\rm ch}_2(V_j)+
             {\rm ch}_2(V_i)\, c_1(V_j).
\eea                
Finally, for the Euler characteristic of the anti-symmetric product bundle 
$\mbox{$\bigwedge^2 V$}$ 
one obtains
\bea
     \chi(\mbox{$\bigwedge^2 V$})=(n-4)\,\chi(V) +
        c_1(V)\left( {\rm ch}_2(V) + {1\over 4} c_2(T)\right)
\eea
and for the symmetric product bundle 
$\mbox{$\bigotimes^2_s V$}$ 
\bea
     \chi(\mbox{$\bigotimes^2_s V$})=(n+4)\,\chi(V) +
        c_1(V)\left( {\rm ch}_2(V) - {1\over 4} c_2(T)\right).
\eea

\subsection{The supersymmetry conditions}

Now we have collected most of  the ingredients to construct bundles and compute the chiral
 spectrum of $SO(32)$ heterotic  string compactifications  on 
elliptically fibered Calabi-Yau threefolds. 
It only remains to evaluate the loop-corrected DUY condition. For  the 
$U(n)$ bundles we obtain 
\bea
 &&  {n\over 2}\, r_\sigma\, \biggl( 2J_B - r_\sigma\, c_1(B)\biggr)\,\left( c_1(\zeta) - q\, c_1(B)\right)+\frac{n q}{2}  J_B^2 \nonumber\\
&=& g_s^2 \left[ \chi(V) -{n\over 2} c_1(\zeta)\, c_1(B) -{n q \over 24}\left(c_2(B)-c_1(B)^2 \right) \right]
\eea
after expressing $J=\ell_s^2 (r_\sigma\, \sigma + J_B)$ in terms of $J_B$, the K\"ahler form on the base $B$. This equation has to be satisfied inside the K\"ahler cone for the model to be well-defined. The constraints on the K\"ahler moduli resulting from this requirement are collected in Appendix A.

The positivity condition on the real part of the gauge kinetic function for a $U(N)$ factor 
leads to the second constraint
\bea
&& {n \over 3!} r_{\sigma} \left( r_{\sigma}^2c_1(B)^2 - 3r_{\sigma} c_1(B)J_B + 3 J_B^2  \right) \nonumber \\
&-& g_s^2 \left[ 
\left( r_{\sigma}c_1(B)-J_B \right)\left( \eta-{n q \over 2} \left(2c_1(\zeta)-qc_1(B)\right)\right) 
+r_{\sigma}a_F \right] \nonumber \\
&-& g_s^2 \frac{n}{2} \left[ c_1(B) J_B +\frac{r_{\sigma}}{12}\left(c_2(B)-c_1(B)^2 \right)
\right] >0.
\eea

These conditions impose strong constraints on the 
bundles to be put simultaneously on the manifold $X$.
In general each $U(n)$ bundle freezes
one combination of the dilaton and the $b_2(B)+1$ radii.

\section{Semi-realistic models on del Pezzo surfaces}
\label{secdelPezzo}

Let us now choose as the base manifold  one of the
 del Pezzo surfaces dP$_r$ with $r=0,\ldots,9$,  which are defined by blowing up
$r$ points on $\IP_2$. This means that $H^2(\rm{dP}_r)$ is generated by the $r+1$ 
elements $l,E_1,\ldots,E_r$, where the class $l$ is inherited from 
$\IP_2$ and the $E_m$ denote the $r$  exceptional cycles introduced
by the blow-ups. 
The intersection form can be computed as
\bea
\label{Intform}
      l\cdot l=1, \quad l\cdot E_m=0, \quad E_m\cdot E_n=-\delta_{m,n},
\eea
and the Chern classes read
\bea
   c_1(dP_r)=3l-\sum_{m=1}^r  E_m, \quad\quad c_2(dP_r)=3+r.
\eea
For the second Chern class of $X$ we obtain
\bea
   c_2(TX)=12 \sigma c_1(B) +(102-10r)\, F.
\eea
Now for a vector bundle $V_i$ we can expand $\eta_i$ and $c_1(\zeta_i)$ in a cohomological basis
\bea
   \eta_i=\eta_i^{(0)}\, l + \sum_{m=1}^r \eta_i^{(m)}\, E_m, \quad\quad\quad
   c_1(\zeta_i)=\zeta_i^{(0)}\, l + \sum_{m=1}^r  \zeta_i^{(m)}\, E_m.
\eea    
As mentioned before we have to require that $\eta$ is effective
and that for stability $\eta-n\, c_1(B)$ is effective
as well. Fortunately, all effective curves on dP$_r$ have been
classified in \cite{Demazure} and we list the reformulated result of~\cite{Donagi:2004ia}
in Table~\ref{TdPrGenerators} for
completeness.
\begin{table}[htb]
\renewcommand{\arraystretch}{1.5}
\begin{center}
\begin{tabular}{|c||c|c|}
\hline
\hline
$r$ & Generators & $\#$  \\
\hline \hline
1 & $E_1$, $l-E_1$  & 2 \\\hline
2 & $E_i$, $l-E_1-E_2$  & 3 \\\hline
3 & $E_i$, $l-E_i-E_j$  & 6 \\\hline  
4 & $E_i$, $l-E_i-E_j$  & 10 \\\hline  
5 & $E_i$, $l-E_i-E_j$, $2l-E_1-E_2-E_3-E_4-E_5$ & 16  \\\hline    
6 & $E_i$, $l-E_i-E_j$, $2l-E_i-E_j-E_k-E_l-E_m$ & 27  \\\hline        
7 & $E_i$, $l-E_i-E_j$, $2l-E_i-E_j-E_k-E_l-E_m$, & \\
 &  $3l-2E_i-E_j-E_k-E_l-E_m-E_n-E_o$ & 56  \\\hline       
8 & $E_i$, $l-E_i-E_j$, $2l-E_i-E_j-E_k-E_l-E_m$, & \\
 &  $3l-2E_i-E_j-E_k-E_l-E_m-E_n-E_o$, & \\  
 & $4l-2(E_i+E_j+E_k)-\sum_{i=1}^5 E_{m_i}$, & \\
 & $5l-2\sum_{i=1}^6E_{m_i}-E_k-E_l$, $6l-3E_i-2 \sum_{i=1}^7 E_{m_i}$ & 240  \\\hline   
9 & $f=3-\sum_{i=1}^9 E_i$, and $\{y_a\}$ with $y_a^2=-1$, $y_a \cdot f=1$ & $\infty$
\\
\hline
\end{tabular}
\caption{\small Generators for the Mori cone of each dP$_r$, $r=1,\ldots,9$. All indices $i,j,\ldots \in \{1,\ldots,r\}$ 
in the table are distinct. The effective classes can be written as linear combinations of the generators with integer 
non-negative coefficients. }
\label{TdPrGenerators}
\end{center}
\end{table}

Moreover, $|\eta|$ is known to be base point free if $\eta\cdot E\ge 0$
for every curve $E$ with $E^2=-1$ and $E\cdot c_1(B)=1$. Such curves
are precisely given by the generators of the Mori cone listed
in Table~\ref{TdPrGenerators}.

From our discussion it is clear that the parameter space of potentially consistent vacua is extremely huge and a completely systematic search for interesting models appears challenging. In the remainder of this section we present two semi-realistic examples which our very preliminary and restrictive survey has produced and whose properties are typical of a large set of solutions that can easily be generated. In fact, we have only covered a tiny fraction of the solution space of vector bundles on elliptic fibrations over dP$_3$ and dP$_4$.

\subsection{ A four-generation Pati-Salam model on dP$_3$ }

As a first example we choose the basis of the elliptic fibration to
be the del Pezzo surface dP$_3$.
Then  we embed a bundle with structure group $U(1) \times U(2)^2$ into $U(4)^3$
yielding the observable group 
\bea
         H=U(4)\times U(2)^2 \times SO(8).
\eea
The data for the twisted bundles are given in Table~\ref{exaa}.
\begin{table}[htb]
\renewcommand{\arraystretch}{1.5}
\begin{center}
\begin{tabular}{|c|c|c||c|c|}
\hline
\hline
$U(n_i)$  & $\lambda_i$ & $\eta_i$ & $q_i$  & $\zeta_i$  \\
\hline
\hline
$U(1)_a$ & 0 & 0 &  0 &  $-2l + 3 E_2 +3E_3$ \\
\hline
$U(2)_b$ & $0$ &  $11 l -3E_1-5E_2-3E_3$ &  $0$ &  $-l+\sum_{m=1}^3 E_m$ \\  
\hline
$U(2)_c$ & $0$ & $7l-3E_1-E_2-3E_3$ & $0$ & $-4l+4\sum_{m=1}^3 E_m$ \\
\hline
\end{tabular}
\caption{\small Defining data for a $U(1) \times U(2)^2$ bundle.}
\label{exaa}
\end{center}
\end{table}

It can be checked explicitly that $\eta_b+c_1(B)$ and $\eta_c+c_1(B)$ 
have only even expansion coefficients as required for $n$ even and $\lambda \in \mathbb{Z}$. 
Furthermore, $\eta_b$ and $\eta_c$ as well as
\bea
\eta_b - 2 c_1(B) = 5l-E_1-3E_2-E_3,\quad
\eta_c - 2 c_1(B) = l-E_1+E_2-E_3
\eea
are effective and the linear systems $|\eta_b|$, $|\eta_c|$ are base point free,
i.e. all intersections with the basis of the Mori cone listed in Table~\ref{TdPrGenerators}
are non-negative. Therefore, the constructed bundles are indeed $\mu$-stable.

Finally, the tadpole 
\bea
c_2(T) = 12\left[3l-\sum_{m=1}^3 E_m \right]\sigma+72
\eea
is cancelled without adding H5-branes due to
\bea
{\rm ch}_2(V_a) &=& -7, \nonumber\\
{\rm ch}_2(V_b) &=& \left[-11 l +3 E_1 +5 E_2 +3 E_3  \right] \sigma + 8, \nonumber\\
{\rm ch}_2(V_c) &=&  \left[-7l+3E_1+E_2+3E_3  \right]\sigma - 30.
\eea
The resulting chiral spectrum is displayed in Table~\ref{T4GenPS}.
Observe in particular that  there is no chiral state charged under $SO(8)$ due to $\chi(V_i) = 0$
and that there are no symmetric or antisymmetric chiral states since in addition 
$\zeta_i \cdot {\rm ch}_2(V_i)=\zeta_i \cdot c_2(T)=0$ for all $i$.
 
The analysis of the chiral spectrum shows that all three $U(1)$ factors are  anomaly-free. 
However, the mass matrix (see Appendix B) has rank two, and only the linear combination 
$4U(1)_b-U(1)_c$ remains massless.

\begin{table}[htb]
\renewcommand{\arraystretch}{1.5}
\begin{center}
\begin{tabular}{|c||c|}
\hline
\hline
$U(4)_a \times U(2)_b \times U(2)_c$ & mult.    \\
\hline \hline
$(\ov{\4},\ov{\2},\1)_{-1,-1,0}$ & 2  \\
$(\ov{\4},\2,\1)_{-1,1,0}$ & 2  \\\hline
$(\4,\1,\ov{\2})_{1,0,-1}$ & 2  \\
$(\4,\1,\2)_{1,0,1}$ & 2 \\\hline
\end{tabular}
\caption{\small Chiral spectrum of a four generation Pati-Salam model on dP$_3$. }
\label{T4GenPS}
\end{center}
\end{table}

The resulting DUY conditions are very simple in this configuration since all one-loop contributions
cancel,
\bea
r_{\sigma}\left( 3r_2+3r_3+2r_0 \right) &=& 0, \nonumber\\
r_{\sigma}\left(r_0+\sum_{m=1}^3 r_m\right)  &=& 0, \nonumber
\eea
and the positivity of the gauge kinetic functions requires
\bea
r_{\sigma}\left(2r_{\sigma}^2-r_{\sigma}(3r_0+\sum_{m=1}^3 r_m)+r_0^2 -\sum_{m=1}^3 r_m^2\right) 
-g_s^2 \left(-14r_{\sigma}+ 3r_0+\sum_{m=1}^3r_m \right) &>& 0, \nonumber\\ 
r_{\sigma}\left(2r_{\sigma}^2-r_{\sigma}(3r_0+\sum_{m=1}^3r_m)+r_0^2 -\sum_{m=1}^3r_m^2 \right) 
-g_s^2 \left(30r_{\sigma}-8r_0-2r_1-4r_2-2r_3  \right) &>& 0, \nonumber\\ 
r_{\sigma}\left(2r_{\sigma}^2-r_{\sigma}(3r_0+\sum_{m=1}^3r_m)+r_0^2 -\sum_{m=1}^3r_m^2 \right) 
+g_s^2 \left(16r_{\sigma} + 4r_0+2r_1+2r_3 \right) &>& 0. \nonumber
\eea
These conditions can be fulfilled in the perturbative regime
inside the K\"ahler cone, e.g. for arbitrary $r_{\sigma}$ and 
 $g_s<0.16\, r_{\sigma}$,  $r_0=1.8\, r_{\sigma} $, $r_1=r_2=r_3=-0.6\, r_{\sigma}$.

\subsection{ A three-generation Standard-like model  on dP$_4$ }
\label{MSSMmodel}

This section is devoted to a three-generation Standard-like model involving four vector bundles, where we now take the base manifold to be dP$_4$. 
It can  be regarded as the generalized S-dual version of the four-stack models which have become popular in the framework of intersecting branes.
Our aim is therefore to obtain  a visible gauge group $U(3)_a\times U(2)_b \times U(1)_c \times U(1)_d$ and realize the quarks and leptons as appropriate bifundamentals. A possible  choice of the hypercharge as a (massless) combination of the abelian factors is  given by  $Q_Y= \frac{1}{6} Q_a + \frac{1}{2}(Q_c + Q_d)$. In this case, also some of the  (anti-)symmetric representations carry MSSM quantum numbers . The details of the chiral MSSM spectrum we try to reproduce can be found in Table~\ref{MSSM}.

Among the many possibilities we consider the simple embedding of the structure group $G=U(1) \times U(1) \times U(2) \times U(1)$ into $U(3) \times U(2) \times U(2) \times U(1)$. This leads to
\bea
H=U(3) \times U(2) \times U(1) \times U(1) \times SO(16)
\eea
modulo the issue of anomalous abelian factors. We choose the bundles
characterized in Table~\ref{exab}.

\begin{table}[htb]
\renewcommand{\arraystretch}{1.5}
\begin{center}
\begin{tabular}{|c|c|c||c|c|}
\hline
\hline
$U(n_i)$  & $\lambda_i$ & $\eta_i$ & $q_i$  & $\zeta_i$  \\
\hline
\hline
$U(1)_a$ & 0 & 0  & 1 &  $5l-3 E_1 - 5 E_2 - E_3    $ \\
\hline
$U(1)_b$ &  0 & 0  & 1  & $-3l +5E_1 +2 E_2 - E_3 + E_4$  \\  
\hline
$SU(2)_c$ & $0$ & $7l-3E_1-3E_2-E_3-E_4$  & $0$ & 0  \\
\hline
$U(1)_d$ & 0  & 0  & - 1 &  $- 5l+3 E_1 +5 E_2 + E_3    $ \\
\hline
\end{tabular}
\caption{\small Defining data for a $U(1) \times U(1) \times SU(2) \times U(1)$ bundle.}
\label{exab}
\end{center}
\end{table}

Note that $V_c$ actually has  structure group $SU(2)$ rather than $U(2)$ since its first Chern class vanishes, which however 
makes no difference in the group theoretic decomposition of $SO(32)$.
Again, one may verify explicitly that the conditions for $\mu$-stability are satisfied. Let us also point out that the 
requirement~(\ref{K}) of cancellation of the Witten anomaly, which is non-trivial for odd $N_a$, is satisfied by the 
configuration. Furthermore, the $U(1)_Y$ hypercharge is indeed massless as desired (see Appendix B). However, since the rank of the 
mass matrix is two, we get another massless $U(1)$ in the four-dimensional gauge group, which is readily identified as $U(1)_c$. 
The perturbative low energy gauge group is therefore
\bea
H=SU(3) \times SU(2) \times U(1)_Y \times U(1)'\times SO(16).
\eea

The degeneracy of the bundle $V_a$ and $V_d= V_a^*$  leads to a gauge enhancement of the $U(3)_a$ and the $U(1)_d$ to 
a $U(4)$. 
Apart from these drawbacks, the
configuration indeed gives rise to three families of the MSSM chiral
spectrum as listed in Table~\ref{MSSM}.   

\begin{table}[htb]
\renewcommand{\arraystretch}{1.5}
\begin{center}
\begin{tabular}{|c||c||c||c||c||c}
\hline
\hline
\multicolumn{5}{|c|}{$U(3)_a\times U(2)_b \times U(1)_c \times U(1)_d \times SO(16) \times \prod_a Sp(2N_a)$}\\\hline
MSSM particle & repr. & index   & mult. & total  \\\hline
$Q_L$ & $(\3,\ov{\2};1,1)_{(1,-1,0,0)}$   & $ \chi(X, V_a \otimes V_b^*)  $ & 8 &  \\
$Q_L$ & $(\3,\2;1,1)_{(1,1,0,0)}$   & $\chi(X, V_a \otimes V_b)      $  &  -11 & -3\\\hline
$u_R$ &  $(\ov{\3},1;1,1)_{(-1,0,-1,0)}$  &  $\chi(X, V_a^* \otimes V_c^*)$ & -3 & \\
$u_R$ &  $(\ov{\3},1;1,1)_{(-1,0,0,-1)}$  &  $\chi(X, V_a^* \otimes V_d^*)$ & 0  & -3 \\ \hline
$d_R$ &  $(\ov{\3},1;1,1)_{(-1,0,1,0)}$  &  $\chi(X, V_a^* \otimes V_c) $ & -3  & \\
$d_R$ &  $(\ov{\3},1;1,1)_{(-1,0,0,1)}$  &  $\chi(X, V_a^* \otimes V_d) $ & 45  &  \\  
$d_R$ & $(\ov{\3}_A,1;1,1)_{(2,0,0,0)}$  &  $  \chi(X,\bigotimes_s^2 V_a)  $ & -45 & -3\\ \hline
$L$ & $(1,\2;1,1)_{(0,1,-1,0)}$ & $ \chi(X, V_b \otimes V_c^*)$ &  -7 & \\ 
$L$ & $(1,\2;1,1)_{(0,1,0,-1)}$ & $ \chi(X, V_b \otimes V_d^*)$ &  -11 &\\  
$L$ & $(1,\ov{\2};1,1)_{(0,-1,-1,0)}$ & $ \chi(X, V_b^* \otimes V_c^*)$ & 7 &   \\ 
$L$ & $(1,\ov{\2};1,1)_{(0,-1,0,-1)}$ & $ \chi(X, V_b^* \otimes V_d^*)$ & 8 & -3 \\ \hline
$e_R$ & $(1,1;1,1)_{(0,0,2,0)}$ & $  \chi(X,\bigwedge^2 V_c)$  & 0  &     \\
$e_R$ & $(1,1;1,1)_{(0,0,0,2)}$ & $ \chi(X,\bigwedge^2 V_d) $ &  0  &\\
$e_R$ & $(1,1;1,1)_{(0,0,1,1)}$ & $\chi(X, V_c \otimes V_d) $ &  -3 & -3 \\ \hline 
$\nu_R$ & $(1,1;1,1)_{(0,0,-1,1)}$ & $\chi(X, V_c^* \otimes V_d) $ &  -3 & -3 \\
\hline \hline
\end{tabular}
\caption{\small Chiral MSSM spectrum for a four-stack model with $Q_Y= \frac{1}{6}
  Q_a + \frac{1}{2}(Q_c + Q_d)$. 
}  
\label{MSSM}
\end{center}
\end{table}

In addition, we get some chiral  exotic matter in the antisymmetric of 
the $U(2)$ and in the bifundamental of the $SO(16)$ with the $U(3)$ and $U(2)$, respectively (see Table~\ref{MSSMb}).   

\begin{table}[htb]
\renewcommand{\arraystretch}{1.5}
\begin{center}
\begin{tabular}{|c||c||c||c||c||c}
\hline
\hline
\multicolumn{5}{|c|}{$U(3)_a\times U(2)_b \times U(1)_c \times U(1)_d \times SO(16) \times \prod_a Sp(2N_a)$}\\\hline
MSSM particle & repr. & index   & mult. & total  \\\hline
- & $ (1,\1_A;1,1)_{(0,2,0,0)} $& $\chi(X,\bigotimes_s^2 V_b)$  & -77 & -77 \\\hline 
- & $ (\3,1;{\bf 16},1)_{(1,0,0,0)}$ & $\chi(X, V_a)$ & -1& -1  \\ \hline
- & $ (1,\2;{\bf 16},1)_{(0,1,0,0)}$ & $\chi(X, V_b)$ & -11& -11 \\ \hline
- & $ (1,1;{\bf 16},1)_{(0,0,1,0)}$ & $\chi(X, V_c)$ &  0  &  0 \\ \hline
- & $ (1,1;{\bf 16},1)_{(0,0,0,1)}$ & $\chi(X, V_d)$ &  1 &  1 \\ \hline 
- & $\sum_a (\3,1;1, {\bf 2N_a})_{(1,0,0,0)}$ & $ \chi(X, V_a{\otimes \cal O}\vert_{\Gamma} )$ & 8& 8 \\
- & $\sum_a (1,\2;1, {\bf 2N_a})_{(0,1,0,0)}$ & $ \chi(X, V_b{\otimes \cal O}\vert_{\Gamma} )$ & 56 & 56 \\
- & $\sum_a (1,1;1, {\bf 2N_a})_{(0,0,1,0)}$ & $ \chi(X, V_c{\otimes \cal O}\vert_{\Gamma} )$ & 0& 0 \\
- & $\sum_a (1,1;1, {\bf 2N_a})_{(0,0,0,1)}$ & $ \chi(X, V_d{\otimes \cal O}\vert_{\Gamma} )$ & -8& -8 \\\hline \hline 
\end{tabular}
\caption{\small Chiral exotic spectrum for the four-stack model with $Q_Y= \frac{1}{6} Q_a + \frac{1}{2}(Q_c + Q_d)$.
In the second column, the first two entries refer to the $U(3)$ and $U(2)$ factors, 
the  third to the SO(16) group and the fourth collectively represents the symplectic charges. 
The $U(1)$ charges are read off from the lower-case entries.}  
\label{MSSMb}
\end{center}
\end{table}

In contrast to the previous example, the chosen bundles alone do not satisfy the tadpole cancellation condition. However, 
the resulting tadpole can be cancelled by including H5-branes, which demonstrates the importance of allowing for these non-perturbative objects.
From the general form of the tadpole equation we find the four-form characterizing this tadpole to be
\bea
[W]= c_2(T) + \sum_{i=1}^4  N_i\, {\rm  ch}_2(V_i) = 22 \,F + (34 l -8 E_1 -22 E_2 -14 E_3 -6 E_4) \, \sigma.
\eea

Its  Poincar\'e dual  class $[\Gamma] = 22 \sigma + 34 l -8 E_1 -22 E_2 -14 E_3 -6 E_4$ lies inside the Mori cone, 
i.e. is effective, and can thus be regarded as the homology class associated to a (reducible) holomorphic curve 
around which we may wrap a system of  H5-branes. To determine the detailed spectrum and gauge group supported by 
the branes we must choose a decomposition of $[\Gamma]$ into irreducible effective classes around each of which 
we can wrap one H5-brane. These  are given precisely by the generators of the Mori cone in Table \ref{TdPrGenerators}. 
Note that the decomposition is not unique and constitutes (part of) the moduli space of our model; what is universal 
is the total number of chiral degrees of freedom charged under the symplectic sector (see Table \ref{MSSMb}) 
and its total rank. In our case, the latter is easily found to be 74.   
For instance,  the decomposition
\bea
[\Gamma] =22 \, \sigma +  22 ( l - E_2 -E_3) + 12 (l-E_1 - E_4) + 4 E_1 + 8 E_3  + 6 E_4  
\eea
 results in the symplectic gauge group $Sp(44) \times Sp(44)\times Sp(24) \times Sp(8) \times Sp(16) \times Sp(12)$. The bifundamental exotics between the MSSM group and this symplectic gauge sector can be determined with the help of (\ref{ind}). Ideally, this group would be hidden, of course.

The only independent DUY equations are those for $V_a$ and $V_b$
\bea
\frac{1}{2}(r_0^2 - \sum_{m=1}^4 r_m^2) + r_{\sigma}(2 r_0 + 2 r_1 + 4 r_2 - r_4 - \frac{1}{2} r_{\sigma}) &=& -  \frac{49}{12} g_s^2,  \\ 
\frac{1}{2}(r_0^2 - \sum_{m=1}^4 r_m^2) + r_{\sigma}(-6  r_0  -6  r_1 - 3  r_2 -2  r_4 + \frac{7}{2} r_{\sigma}) &=& -  \frac{121}{12} g_s^2,
\eea
and only fix two of the K\"ahler moduli. This can well be achieved, together with positivity of the real part of the various gauge kinetic functions, inside the  K\"ahler cone (see Appendix A) and in the perturbative regime. E.g. by taking  $  r_2= -2.5\,  r_{\sigma}, \, r_3=-1.1\,  r_{\sigma}, \,  r_4= -\,  r_{\sigma}$ and $g_s< 0.58\,  r_{\sigma}$ for arbitrary $\,  r_{\sigma}$, the solution for $r_0$ and $r_1$ satisfies all K\"ahler cone constraints. We can therefore always choose $r_{\sigma}$ and $g_s$ such that the model is indeed in the perturbative regime.

\section{Conclusions}
\label{secConclusions}

In this paper we have continued the systematic analysis  of compactifications
of the $SO(32)$ heterotic string on Calabi-Yau manifolds
\footnote{In a recent paper \cite{Polchinski:2005bg} it was pointed out that, in constrast
to the $E_8\times E_8$ heterotic string, the $SO(32)$ heterotic string
theory might also contain non-perturbative open strings.}. Generalizing
our earlier work \cite{Blumenhagen:2005pm} we have allowed for more general embeddings of the
$U(n)$ structure groups giving rise to a reduction of the rank of the
gauge group as well as to chiral fermions transforming in the
symmetric representation of the observable unitary gauge groups. 
In addition we have also included heterotic 5-branes, which contribute
both to the 6-form tadpole and, unlike in the $E_8\times E_8$
heterotic string, to the chiral massless spectrum.

As we demonstrate in \cite{bhwnew},  S-duality relates these heterotic
models to Type I compactifications, where of course
the orders of terms in string and sigma model perturbation theory are
different. The topological sector, however, is in complete agreement.

It is clear that this framework is extremely rich as long as one has
control of stable vector bundles on certain Calabi-Yau spaces.
Fortunately, on elliptically fibered Calabi-Yau manifolds the
spectral cover construction of FMW 
provides a well controllable set of such stable vector bundles.
In this paper we have considered stable $U(n)$ bundles on elliptically
fibered Calabi-Yau manifolds defined via twisting $SU(n)$ bundles
\`{a} la FMW by line bundles. 

We have only started a systematic search for Standard- respectively GUT-like
models on elliptically fibered Calabi-Yau manifolds, choosing
del Pezzo surfaces for the base.  This paper reports on the very first
results of such a search. Clearly, the models presented
are not yet completely satisfactory, but they demonstrate that not only
the $E_8\times E_8$ heterotic string but also the long
neglected  $SO(32)$ heterotic string can naturally give rise
to Standard model-like string compactifications. The parameter space of
consistent models is so huge that we have restricted our search to only a very
tiny subset of possible solutions, in which we have nonetheless found various
additional three-generation MSSM-like models similar to the one presented in
section \ref{MSSMmodel}. Compared to the tremendous effort that has gone into
the study of toroidal orientifold models and their by far smaller space of
supersymmetric solutions we consider this result quite promising, both from
the point of view of concrete model building and with regard 
to classification attempts in the context of the (open) string landscape.   
A more thorough and systematic search might be carried out
in the context of a future  StringVacuumProject (SVP). 

A nice feature of a certain class of $SU(n)$ bundles on Calabi-Yau 
manifolds is that also the string world-sheet theory
is known to be well behaved and that the vacuum is not destabilised by world-sheet
instanton corrections. This is not so clear for the
class of models discussed here where the left-moving
$U(1)_L$ symmetry on the world-sheet is anomalous. More work is needed to clarify
such more complicated issues. 

\vskip 1cm
 {\noindent  {\large \bf Acknowledgements}}
 \vskip 0.5cm 
It is a great pleasure to thank Bj\"orn Andreas and Florian Gmeiner for helpful discussions and Hakon Enger for 
enlightening correspondence.
 \vskip 2cm

\appendix
\section{K\"ahler cone constraints on Calabi-Yau's with base  dP$_r$}
\label{Kcone}

The DUY equations have to admit solutions for the K\"ahler parameters inside the K\"ahler cone, i.e. such that the integral of powers of the K\"ahler form over all appropriate cycles are positive,
\bea
\int_{2-cycle} J >0 ,  \quad\quad  \int_{4-cycle} J^2 >0 , \quad\quad \int_X J^3>0.
\eea
We expand the K\"ahler form on the elliptically fibered Calabi-Yau  as $J=l_s^2\, (r_{\sigma} \, \sigma + J_B)$ with $J_B= r_0 \, l + \sum_{m=1}^r r_m E_m$ being the  K\"ahler form on the base manifold dP$_r$ in terms of the canonical basis.

From the first constraint we read immediately that the radii must satisfy

\bea
r_{\sigma} >0, \quad\quad    r_0 >0, \quad\quad r_m <0 \ \    {\mbox{for}} \ m   \in \{1,...,r\}.
\eea
The second inequality, $\int J^2 >0$,  holds precisely if in addition

\bea 
r_0^2 -  \sum_{m=1}^r r_m^2 >0, \quad  r_{\sigma}< \frac{2}{3}\, r_0, \quad  r_{\sigma}< -2 r_m  \quad
 {\mbox {for}} \   m \in \{1, .., r\}.
\eea

Finally positivity of the volume of the Calabi-Yau necessitates that also

\bea
r_{\sigma}^3 \,(9-r) - 3 r_{\sigma}^2 \, (3 r_0 + \sum_{m=1}^r r_m) + 3 r_{\sigma} \, (r_0^2 -  \sum_{m=1}^r r_m^2) >0.
\eea

\section{Details of the generalized Green-Schwarz mechanism }
\label{AppTrace}

The trace identities of~\cite{Blumenhagen:2005pm} are generalized for $U(n_i)$ factors diagonally embedded into $U(n_iN_i)$ to
\bea
 {\rm Tr} F{\ov F}^3 &=& 12\sum_{j=1}^{K}N_j \; f_j \wedge \left( 4 {\rm tr}_{U(n_j)} {\ov {F}}^3 
+  {\rm tr}_{U(n_j)} {\ov {F}}   \sum_{i=1}^{K} N_i {\rm tr}_{U(n_i)} {\ov {F}}^2\right), \nonumber\\
{\rm Tr} F^2 {\ov F}^2 &=&
4 \sum_{j=1}^{K} \left( {\rm tr}_{SU(N_j)} F^2 +N_j\; (f_j)^2   \right)
\wedge \Bigl( 12\; {\rm tr}_{U(n_j)} {\ov {F}}^2  + n_j \sum_{i=1}^{K}N_i \; {\rm tr}_{U(n_i)} {\ov {F}}^2 \Bigr) \nonumber \\
&+&8\sum_{i,j=1}^{K} N_i N_j \;\, f_i \,  f_j \wedge  {\rm tr}_{U(n_i)}{\ov {F}}  \,  {\rm tr}_{U(n_j)}{\ov {F}}
+\; 2 \; {\rm tr}_{SO(2M)} F^2 \wedge \sum_{j=1}^{K} N_j \; {\rm tr}_{U(n_j)} {\ov {F}}^2   , \nonumber\\
{\rm Tr} F^2&=& 30 \, {\rm tr}_{SO(2M)} F^2 + 60 \sum_{j=1}^{K} n_j \left( {\rm tr}_{SU(N_j)}{F}^2 +N_j (f_j)^2\right)    , \nonumber\\
{\rm Tr}F {\ov F} &=& 60 \sum_{j=1}^{K} N_j f_j\wedge {\rm tr}_{U(n_j)}{ \ov{F}},  \nonumber\\ 
{\rm Tr} {\ov F}^2 &=& 60 \sum_{j=1}^{K} N_j  \;{\rm tr}_{U(n_j)} {\ov {F}}^2  .
\eea
For details of the notation see section 3 and Appendix B in~\cite{Blumenhagen:2005pm}.

With these relations, the mass terms for the axions are given by
\bea
S^{1-loop}_{mass}
&=&  \frac{ 1 }{6 (2 \pi)^5 \alpha'} \sum_{i=1}^{K}  N_i \int_{{\mathbb
    R}_{1,3}} b_0^{(2)} \wedge  f_i \;  \int_X \Bigl( {\rm tr}_{U(n_i)} {\ov {F}}^3 - \frac{1}{16} 
 {\rm tr}_{U(n_i)}{\ov {F}} \wedge  {\rm tr}{\ov R}^2  \Bigr), \nonumber\\
S^{tree}_{mass} &=& \frac{1}{ (2 \pi)^2
  \alpha'}\sum_{i=1}^{K}\sum_{k=1}^{h_{11}} N_i\, 
\int_{{\mathbb R}_{1,3}}(  b_k^{(2)} \wedge  f_i)\;  [{\rm tr}_{U(n_i)}{\ov {F}}]_k.
\eea
and the vertex couplings to the gauge fields and gravitons are given by
\bea
S^{1-loop}_{vertex} &=& 
\frac{1}{ 2 \pi} \sum_{k=1}^{h_{11}}\int_{{\mathbb R}_{1,3}} b_k^{(0)}\,\wedge 
\Bigl\{
\sum_{i=1}^K \left({\rm tr}_{SU(N_i)} F^2 + N_i (f_i)^2  \right)
\Bigl[ \frac{1}{4} \,{\rm tr}_{U(n_i)}{\ov {F}^2}   - \frac{n_i}{192} {\rm tr} {\ov R}^2 \Bigr]_k 
\nonumber\\
&&\quad\quad\quad
-\frac{1}{384}{\rm tr}_{SO(2M)} F^2 \; [{\rm tr} {\ov R}^2]_k 
+ \frac{1}{768} {\rm tr} R^2 \Bigl[{\rm tr} {\ov R}^2-4\sum_{i=1}^K N_i {\rm tr}_{U(n_i)}{\ov {F}^2} \Bigr]_k \Bigr\}, \nonumber\\
S^{tree}_{vertex} &=& \frac{1}{8 \pi} \int_{{\mathbb R}_{1,3}} b_0^{(0)} \wedge 
 \Bigl(2\sum_{i=1}^K n_i\left({\rm tr}_{SU(N_i)} F^2 +N_i (f_i)^2\right)+{\rm tr}_{SO(2M)} F^2 
-{\rm tr} R^2 \Bigr).
\nonumber
\eea
In addition, the H5-brane action~(\ref{CS}) contains the vertex couplings
\bea
S^{H5}_{vertex} &=& -\frac{1}{8\pi} \sum_{k=1}^{h_{11}} \int_{\mathbb R_{1,3}}[\Gamma_a]_k\,\,
 b_k^{(0)}\,\wedge \left(
 {\rm tr}_{Sp(2N_a)} F_a^2-\frac{N_a}{24}\, {\rm tr} R^2\right)
\eea
with $[\Gamma_a]_k=\int_{\Gamma_a} \omega_k$.

One may verify that these couplings precisely cancel all abelian anomalies, in the case of the mixed 
abelian-gravitational ones only upon tadpole cancellation.

The mass terms can be rewritten as
\bea
                 S_{mass}=\sum_{i=1}^{K}  \sum_{k=0}^{h_{11}} 
            {M^i_k\over 2\pi\alpha'}
                \int_{\IR_{1,3}} f_i \wedge b^{(2)}_k 
\eea
with 
\bea
 {M^i_k} = \frac{N_i}{2\pi}\times 
\cases{ \chi(V_i)-\frac{1}{24}\int_X c_1(V_i)c_2(T) &  for $k=0$ \cr
[c_1(V_i)]_k &  for $k\in\{1,\ldots,h_{11}\}$, \cr} 
\eea
where $[c_1(V_i)]_k$ is the expansion coefficient of the $k^{th}$ basic two form.
The number of massive $U(1)$ factors is then given by ${\rm rank}(M)$.

The gauge kinetic functions for the unitary factors are easily read off from the vertex couplings
\bea
{\rm f}_{SU(N_i)} &=& 2n_i\, S +\sum_{k=1}^{h_{11}} T_k\, \Bigl({\rm tr}_{U(n_i)}(\ov F^2)_k-
 \frac{n_i}{48}({\rm tr} \ov R^2)_k \Bigr) ,
\nonumber\\
{\rm f}_{U(1)_i} &=& N_i {\rm f}_{SU(N_i)} 
\eea
generalising the result in~\cite{Blumenhagen:2005pm} to arbitrary $n_i$. 
In addition, the gauge kinetic functions for the symplectic factors are
\bea
{\rm f}_{Sp(2N_a)} &=& \frac{1}{2\pi \ell_s^2}\,\int_{\Gamma_a}\left(J-iB\right).
\eea

For the elliptic fibrations over del Pezzo surfaces discussed in this article, we obtain
\bea
\left(\begin{array}{c} 
M^i_0 \\ M^i_{\sigma} \\  M^i_l \\ M^i_{m:m=1,\ldots,r}
\end{array}\right)
 = \frac{N_i}{2\pi}\times 
\left(\begin{array}{c} 
\chi(V_i)+ n_i \left(q_i \left(\frac{1}{4}-\frac{r}{12}\right) -\frac{1}{2} \left(3\zeta^{(0)}_i +\sum_{m=1}^r \zeta^{(m)}_i\right)\right)\\
n_i q_i \\
n_i\zeta^{(0)}_i \\ 
n_i \zeta^{(m)}_i
\end{array}\right).
\eea

\clearpage
\nocite{*}
\bibliography{SOrev}
\bibliographystyle{utphys}

\end{document}